\documentclass{emulateapj}
\usepackage{float, amsmath,verbatim}
\pdfoutput=1

\shorttitle{Electron-Ion Collisionless Shocks} 
\shortauthors{Spitkovsky}

\begin{document}

\title{On the structure of relativistic collisionless shocks in electron-ion plasmas} 

\author{Anatoly Spitkovsky\altaffilmark{1}}
\altaffiltext{1} {Department of Astrophysical Sciences, Peyton Hall,
Princeton University, Princeton, NJ 08544: anatoly@astro.princeton.edu}

\subjectheadings{acceleration of particles -- collisionless shocks -- gamma rays: bursts }
\begin{abstract}

Relativistic collisionless shocks in electron-ion plasma are thought to occur in the afterglow phase of Gamma-Ray Bursts (GRBs), and in other environments where relativistic flows interact with the interstellar medium. A particular regime of shocks in an unmagnetized plasma has generated much interest for GRB applications. In this paper we present ab-initio particle-in-cell simulations of unmagnetized relativistic electron-ion shocks. Using long-term 2.5-dimensional simulations with ion-electron mass ratios from 16 to 1000 we resolve the shock formation and reach a steady-state shock structure beyond the initial transient. We find that even at high ion-electron mass ratios initially unmagnetized shocks can be effectively mediated by the ion Weibel instability with a typical shock thickness of $\sim 50$ ion skin-depths. Upstream of the shock the interaction with merging ion current filaments heats the electron component, so that the postshock flow achieves near equipartition between the ions and electrons, with electron temperature reaching 50\% of the  ion temperature. This energy exchange helps to explain the large electron energy fraction inferred from GRB afterglow observations.
\end{abstract}

\section{Introduction}
Relativistic collisionless shocks propagating in unmagnetized electron-ion plasmas are an essential ingredient in the theory of GRB afterglows (see Piran 2005 and Waxman 2006 for reviews). These shocks are expected to generate sub-equipartition magnetic fields and accelerate nonthermal particles that are responsible for the observed synchrotron emission. Weibel instability has emerged as the leading mechanism for shock formation in weakly magnetized plasmas (Medvedev and Loeb 1999; Gruzinov and Waxman 1999). This instability can generate small scale magnetic fields in counterstreaming plasmas and can isotropize the flow. Weibel instability was simulated by a number of authors in various contexts (e.g., Nishikawa et al. 2003; Silva et al. 2003; Frederiksen et al. 2004;  Medvedev et al. 2005; Nishikawa et al. 2005). Although all groups observe the initial filamentation, sufficiently large simulations that lead to shock formation have only been done in the case of electron-positron pair shocks (Spitkovsky 2005, further S05; Chang et al. 2007, further CSA07). For the electron-ion plasmas the pioneering simulations of Frederiksen et al. (2004), Hededal et al. (2004) and Hededal (2005) were not large enough to see the complete thermalization of the ions and the shock formation. This led Lyubarsky \& Eichler (2006) to question whether the Weibel instability in the electron-ion plasma is fast enough to generate a shock transition that is thinner than the Larmor radius in the weak ISM field. 
In this paper we demonstrate via large numerical simulations that the ion Weibel instability is indeed very effective at establishing the shock transition in an unmagnetized electron-ion plasma even for realistic mass ratios. We study the shock structure and energetics of the flow and find that 
electron heating in the foreshock is an important catalyst to shock formation via filamentation instability. From simulations we find the energy fraction that is transferred to the post-shock electrons and compare it to the values inferred from GRB afterglow observations. 
In \S{\ref{secsimset} we describe the particle-in-cell simulations of shock formation, and then discuss the shock structure in \S{\ref{secshockstr} and electron heating in \S{\ref{secelheat}}.

\section{Simulation setup} \label{secsimset}

Particle-in-cell (PIC) simulations have proven to be a valuable tool in understanding collisionless plasma physics from first principles (Birdsall and Langdon 1992). 
For our study we use a relativistic electromagnetic PIC code {\it TRISTAN-MP}, which is a parallel descendant of the code {\it TRISTAN} (Buneman 1993). Among the modifications we introduced are improved noise properties of the charge deposition and the suppression of numerical Cherenkov instabilities of ultrarelativistic particles. This
enables the propagation of cold relativistic plasmas through the grid for thousands of plasma times (Spitkovsky and Arons 2007, in prep, hereafter SA07). To simulate shock formation we reflect a relativistically moving cold electron-ion stream from a conducting wall. This is equivalent to colliding two streams of identical plasma head-on, but saves 1/2 of the computational effort. We validated this setup by comparing with a true head-on collision. In hydrodynamics a collision of two plasmas should create a contact discontinuity and two shock waves propagating away from the contact into the ``upstream." The wall in our simulation represents the contact, and we simulate only a moving forward shock. Our simulation is performed in the ``downstream" frame of the contact, not in the shock frame. 
Since the downstream plasma is at rest, no extra boosts are needed to study the spectra. 
This helps to disentangle a relativistically moving system of precursors, shocks and contacts that forms in collisions of jets with stationary plasmas (Frederiksen et al. 2004, Nishikawa et al. 2005).
The characteristic scales of interest in an electron-ion shock are on the order of tens of ion skin depths ($c/\omega_{pi}\equiv(4 \pi e^2 n/\gamma m_i c^2)^{-1/2}$, for a plasma with density $n$ and characteristic ion energy $\gamma m_i c^2$). 
We typically use 10 cells per {\it electron} skin depth, so in order to resolve the ion scales for realistic mass ratios we would require simulations with several thousand cells on the side in 3D and tens of billions of particles. This requires a dedicated large-scale simulation effort. Instead, we have chosen to simulate shock formation in 2D, encouraged by the similarity between 2D and 3D shocks in pair plasma (CSA07, S05). 
The simulation plane includes the direction of streaming and one transverse dimension. The magnetic field is out of the plane (CSA07).
We simulated shocks in an initially cold, unmagnetized plasma moving with Lorentz factor $\gamma_0=15$ for a range of ion-electron mass ratios: $m_i/m_e=16, 30, 100, 500, 1000.$ The transverse box size varied so that the box encompassed $\sim40$ upstream ion skin depths (2048 cells for $m_i/m_e=16, 30$; 4096 cells for $m_i/m_e=100$; 8192 cells for $m_i/m_e=500, 1000$). The longitudinal size expanded with the simulation time up to $10^5$ cells for $m_i/m_e=100$, corresponding to $10^3 c/\omega_{pi}$ (this was our longest run). We typically used 2 particles per  cell per species in the unperturbed upstream, which increased to 6 behind the shock due to shock compression. We have tested a larger number of particles per cell and larger transverse box sizes, but the results did not appreciably change.

\begin{figure} 
\centering
\includegraphics[scale=0.5]{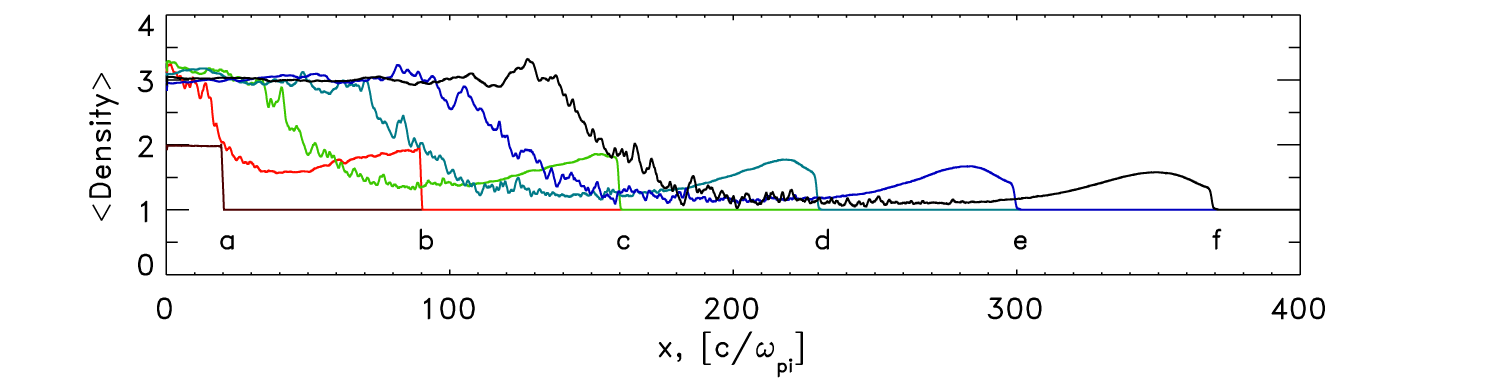}
\caption{Density profiles during the formation of $m_i/m_e=500$ shock. The incoming flow is moving to the left with $\gamma_0=15$. }
\label{figshockstage}
\end{figure}

\section{Shock structure and evolution}\label{secshockstr}
In fig. \ref{figshockstage} we display the time sequence of transversely averaged density profiles that show the formation of a shock in a $m_i/m_e=500$ plasma (the same occurs for other mass ratios). Initially the flow with density $n_1$ and $\gamma_0=15$ is moving to the left. After the bounce from the wall the reflected and the incoming plasmas stream through each other, increasing the density to $2 n_1$ (fig. \ref{figshockstage}a). The electrons undergo Weibel instability and thermalize to their upstream kinetic energy, but the ions are still cold. After $\sim 100 \omega_{pi}^{-1}$ the density near the wall begins to rise, as the ions are randomizing (fig. \ref{figshockstage} b). The ions at the head of the reflected flow are still cold, and propagate close to $c$ into the upstream. This is the ``initialization precursor" -- fast particles that are the remnants of the initial collision. They appear as a moving density bump that always outruns the shock (fig. \ref{figshockstage}b-f). With time this bump is eroded as the particles decelerate. We define the shock as the density compression that propagates away from the wall in fig. \ref{figshockstage}b-f. The shock satisfies the hydrodynamic jump conditions after the initial transient:  $n_2/n_1=\Gamma_{ad}/(\Gamma_{ad}-1) + 1/(\gamma_0 (\Gamma_{ad}-1))=3.13$, where the adiabatic index is $\Gamma_{ad}=3/2$, appropriate for a 2D relativistic gas. 

When the initialization precursor erodes so that the density in front of the shock is close to the unperturbed upstream density, the shock becomes steady. In this stage the integrated quantities through the shock do not significantly evolve as the shock moves though the grid. The shock structure snapshot in this stage is shown in fig. \ref{figshockstr} for a $250 c/\omega_{pi}$ section around the shock from the $m_i/m_e=100$ simulation. The shock speed is close to $v_{shock}=c (\Gamma_{ad}-1) \sqrt{ (\gamma_0-1)/(\gamma_0+1) }=0.47c$, in accord with jump conditions. Plasma density (fig. \ref{figshockstr}a) shows filamentation in the upstream region with filaments reaching $10 c/\omega_{pi}$ in size. Magnetic energy (\ref{figshockstr}b) is also fillamentary with enhancements at the edges of density filaments. Filamentation is driven mainly by the ion dynamics, and the shock compression starts where the filaments merge and disintegrate. The shock transition, which is $\sim50 c/\omega_{pi}$ thick (fig. \ref{figshockstr}c) corresponds to a peak in the magnetic energy  (fig. \ref{figshockstr}d). Incoming flow is isotropized by this magnetic field. While the magnetic energy density in the shock can reach local equipartition with the upstream flow energy ($\epsilon_B \equiv B^2 / 4 \pi \gamma_0 n_1 m_i c^2 \sim1$), the average magnetic energy is near $10\%-15\%$ of equipartition at the shock. Magnetic filaments are destroyed in the shock, and the downstream magnetic field forms islands as in the pair shock simulations (CSA07). The field energy decays below $4\times10^{-3}$ of the upstream energy density; however, runs with more particles/cell are needed to reliably study the downstream field further.  

The filaments of  density and magnetic field are not stationary in front of the shock. They are advected with the upstream flow and merge on the transit time of a filament towards the shock. As a result, the individual clumps that enter the shock change over time; however, the average density profile is relatively stable. 

\begin{figure} 
\centering
\includegraphics[scale=0.54]{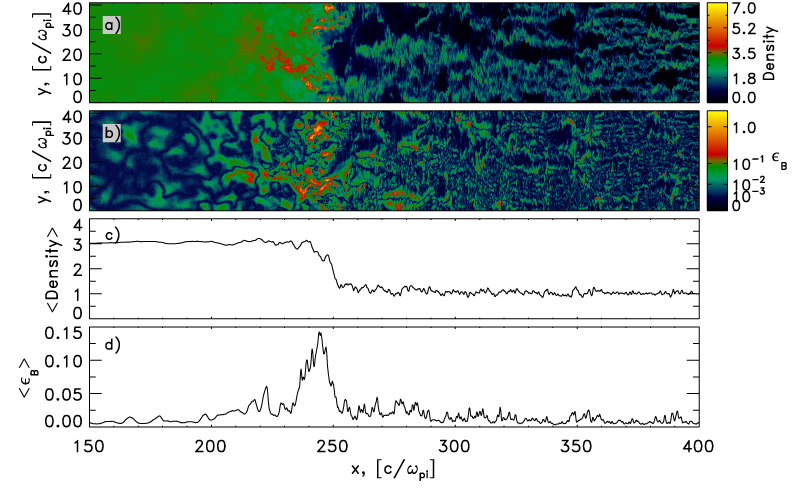}
\caption{Steady state structure of $m_i/m_e=100$ collisionless shock. a) Density structure in simulation plane (normalized by the unperturbed upstream density); b) Magnetic energy density $\epsilon_B$, in units of the upstream energy density; c) Transversely averaged plasma density; d) Transversely averaged  $\epsilon_B$.}
\label{figshockstr}
\end{figure}
\section{Electron Heating}\label{secelheat}
In the steady state the shock is not influenced by the wall or the initial collision. It is a self-propagating structure. 
In order to maintain continuing filamentation in the upstream the cold incoming fluid should experience counterstreaming, which is provided by a population of particles escaping from the downstream (Milosavljevic et al. 2006, SA07). 
In fig. \ref{figshockstrpsp} we show the longitudinal momentum space for ions (fig. \ref{figshockstrpsp}b) and electrons (fig. \ref{figshockstrpsp}c) for $m_i/m_e=100$ run. 
The incoming flow has negative values of four-velocity. In this snapshot the flow is stopped and thermalized at the shock for $x<250 c/\omega_{pi}$ (fig. \ref{figshockstrpsp}a shows the density structure of the shock for reference). There is a clear population of particles with positive four-velocities streaming away from the shock. These particles, though being hotter and more tenuous than the incoming fluid, cause the filamentation instability in the upstream. A structure with positive 4-velocity at $x>450 c/\omega_{pi}$ in fig. \ref{figshockstrpsp}b is the remnant of the initialization precursor. Over time its contribution to the initiation of filamentation diminishes (we have seen it disappear completely in runs with smaller mass ratios). The particles reflected from the shock occupy a region that extends $>300 c/\omega_{pi}$ before the shock for the times to which we have evolved the simulation. 

From fig. \ref{figshockstrpsp}c it is clear that the electrons thermalize with $\gamma$-factors that significantly exceed $\gamma_0$, suggesting electron heating before the shock. To quantify this effect we plot the average energy per particle normalized by the upstream ion energy in fig. \ref{figshockstrpsp}d with red and blue curves for ions and electrons respectively. The solid lines show only the electrons and ions that are moving towards the shock; these lines show the partition of energy in a fluid element as it approaches the shock. The electrons gain 35\% of the initial ion energy by the time they reach the downstream. In the units of downstream ion energy, the electrons are at 50\% of equipartition with the ions. If we include the reflected particles into the mean energy calculation (dashed line in fig. \ref{figshockstrpsp}d), both species show reheating near the shock, indicating that the reflected particles are a separate population that get their energy by bouncing off the shock front and can spend more time near the shock than the transit time of the incoming fluid. The energy spectra of slices from the upstream (fig. \ref{figshockstrpsp}f) and the downstream (fig. \ref{figshockstrpsp}e) show that while in the upstream the ions were a cold beam and electrons were thermalized to their initial energy, in the downstream both species are near Maxwellian distributions with similar temperatures. A non-Maxwellian tail can also be observed, especially in electrons, and is due to reflected particles (Spitkovsky 07, in prep).

We studied the mechanism of electron heating by plotting the orbits of particles from simulations. The ion filaments are not completely neutral, with an excess of positive charge in the middle of the filament and a shielding negative charge on the periphery. Hence, there is an electric field which accelerates electrons to the center of the filament (in addition to giving the $E\times B$ motion with the filament towards the shock). In general, this acceleration is reversible as the electrons lose energy on the way out of the filament (Hededal et al. 2004, Medvedev 2006). However, the development of Weibel instability increases the charge of the filaments with time, so on every return particles experience a deeper potential well and increase the amplitude of energy oscillations. In the nonlinear stage of the instability (near the shock), the filaments merge and break up on the time scale of electron oscillation in the filament. In this region the hot electrons can switch ion filaments while retaining their energy. Electrons are also attracted to isolated clumps of positive charge that result from the break up of ion filaments near the shock. When averaged in the transverse direction these clumps of charge form an attractive electrostatic potential well extending $\sim 200 c/\omega_{pi}$ in front of the shock that accelerates electrons and decelerates ions. The potential is variable on the time scale of transit of charge clumps, so the electrons can retain some of the gained energy after exiting the well and crossing the shock. A detailed theory of the heating process will be presented elsewhere (Medvedev \& Spitkovsky, in prep).  

\begin{figure} 
\centering
\includegraphics[scale=0.47]{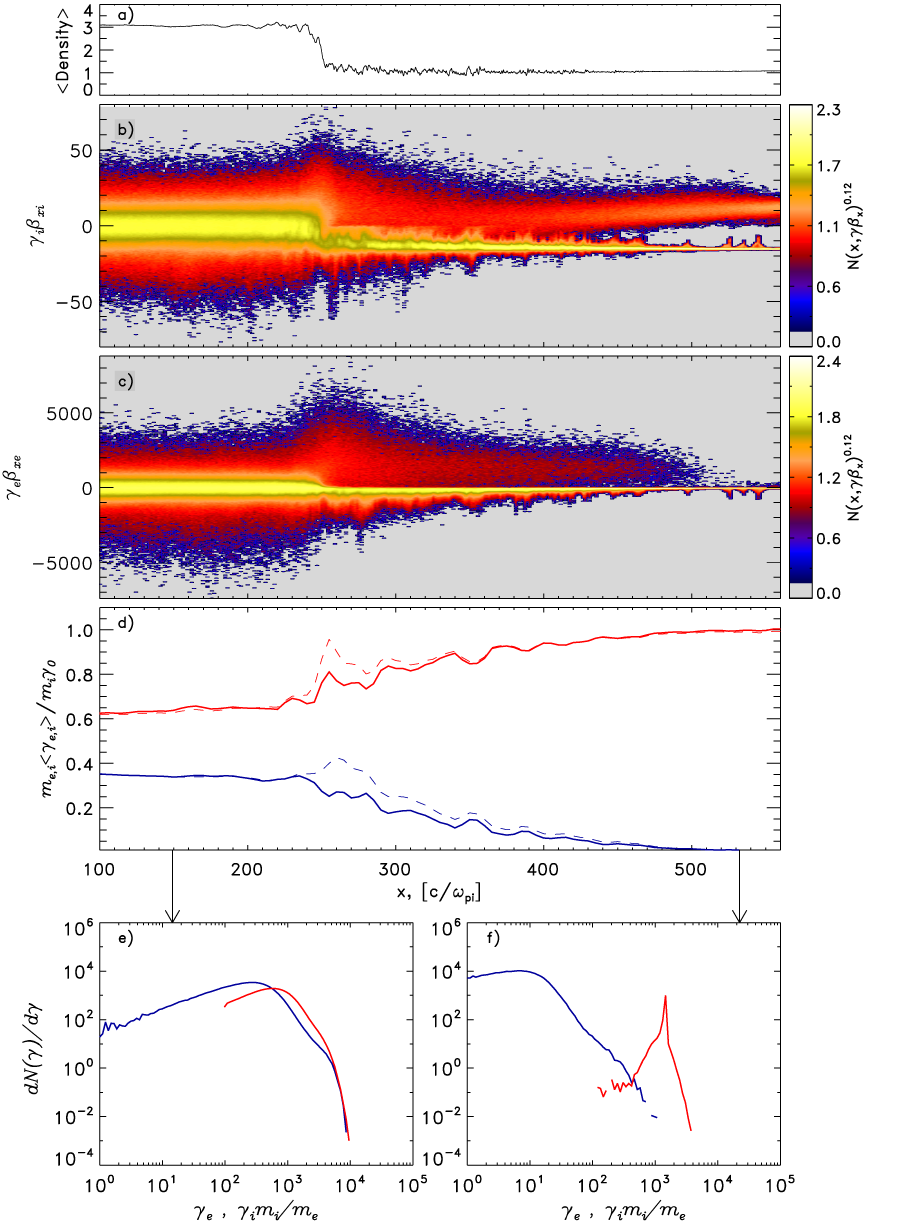}
\vskip -.05in
\caption{Internal structure of $m_i/m_e=100$ shock. a) Average density profile; b) Momentum space density $N(x,\gamma \beta_x)$ for ions; c) Momentum space density for electrons; d) Average particle energy for particles that move towards the shock (in units of upstream ion energy): red -- ions, blue -- electrons. Dashed lines show average energy that includes reflected particles; e) Particle spectrum in the downstream slice centered on $x=150 c/\omega_{pi}$, red for ions, blue for electons; f) Spectrum in the upstream slice at $x=530 c/\omega_{pi}$.}
\label{figshockstrpsp}
\end{figure}

\section{Discussion}\label{discus}

In this paper we studied the structure of collisionless shocks in an initially unmagnetized relativistic electron-ion plasma. Our 2D PIC simulations are sufficiently large to resolve the formation of a shock as a density jump of thickness $\sim 50 c/\omega_{pi}$ which satisfies the hydrodynamic jump conditions. Self-generated transverse magnetic fields reach $15\%$ of the initial ion energy in the shock transition. An important ingredient is the foreshock region that extends several hundred $c/\omega_{pi}$ into the upstream. Here, the incoming cold flow interacts with the hot tenuous plasma that escapes from the downstream.  
This interaction leads to the ion Weibel instability and the formation of filaments of current and density. Electrons oscillating in the electromagnetic field of the growing and merging ion filaments are efficiently heated close to equipartition with the ions. This heating is essential to the formation of unmagnetized shocks. As was pointed out by Lyubarsky \& Eichler (2006), ion Weibel instability slows down if electrons have temperature much smaller than the ion energy. Such electrons can efficiently screen the ion filaments from mutual attraction. Electron heating reduces this shielding, increasing the electron Larmor radius and skin depth, and allowing them to leave the filaments more easily. This increases the charge separation inside the filaments and facilitates merging. 
Electron acceleration in ion filaments was previously discussed by Hededal et al. (2004) and Medvedev (2006). In the shock context we can also point out the effects that can allow electrons to keep the gained energy: the evolution and merging of filaments on the transit time to the shock and the escape of energized electrons from the filaments. This creates a time-varying potential well in front of the shock, which contributes to net heating. 

The amount of heating we find is considerable -- 
the downstream electron energy fraction is $\epsilon_e\sim 0.5$. This supports the canonical picture of GRB afterglows, where electron energy fraction in the radiating electrons $\epsilon_e \sim 0.1$ is commonly inferred (Piran 2005). We note that the bulk of our downstream electrons is thermal (fig. \ref{figshockstrpsp}f), so it is reasonable to assume that the energy fraction in the nonthermal component, once it appears, would be smaller than what we obtain. Our result also constrains the minimum electron Lorentz factor in an accelerated power-law to be $\gtrsim\gamma_0 m_p/m_e$ (cf. Eichler \& Waxman 2005).  The heating mechanism is robust even in the absence of a shock: we performed 2D periodic box simulations of counterstreaming electron-ion plasmas with $\gamma_0=15$ for mass ratios of $100$ and $500$. In both cases the electrons gained $24\%$ of the initial flow energy, which is smaller than in the shock setup, but still much larger than the original electron energy. 
 
We studied the properties of the shocks as a function of electron-ion mass ratio. Interestingly, for $m_i/m_e>30$, the properties such as shock width (in $c/\omega_{pi}$) and magnetic and electron energy fractions do not significantly change with ion mass. Although the largest mass ratio we tried was $1000$, the convergence of shock properties with mass suggests to expect no surprises even at $m_i/m_e=1836$. Also, shocks with $10<\gamma_0<100$ show very similar behavior, so our results for $\gamma_0=15$ are representative of ultra-relativistic shocks.

The shock structure seems to have reached a steady state in our simulations. This means that we do not see a significant evolution in the integrated quantities as the shock propagates through the grid on time scales of our runs (up to $10^3 \omega_{pi}^{-1}$). However, there are some features that continue to evolve. There are particles that are escaping into the upstream at $\sim c$, and with time the spatial extent of the ``cloud" of reflected particles increases. If there are Fermi-type processes in the shock, they would be associated with these particles (Spitkovsky 07, in prep). Based on our simulations we cannot rule out the formation of high energy particle population that could change the shock structure on longer timescales. 

A lingering question is whether our 2D simulations capture the relevant 3D physics. To check this we performed a $m_i/m_e=16$ simulation of a 3D relativistic shock on a relatively small $256^2 \times 5000$ grid (only $6^2 \times 125 c/\omega_{pi}$). We obtain the density ramp and the formation of a shock with electron heating at $30\%$ of the upstream energy that is very similar to the 2D case. From this we expect that our 2D results for higher mass ratios will hold in 3D. 

Electron heating to near-equipartition with the ions in the shock implies that relativistic electron-ion shocks should be similar to the electron-positron pair shocks, especially in the downstream region. There, both electrons and ions have similar effective mass and Larmor radii. The behavior of the downstream magnetic turbulence should then be very similar between the pair and electron-ion shocks. The magnetic islands in the downstream of fig. \ref{figshockstr}b are on the scale $10 c/\omega_{pi}$ (which is also close to $10 c/\omega_{pe}$ in the downstream electron skin depths). Similar islands were observed in CSA07 for the downstream of pair shocks.  We expect that higher resolution simulations of the downstream region in e-ion shocks will find similar decay of the field with time as in CSA07. 

Recently, the interest in understanding the electron-ion temperature ratio in collisionless shocks has increased due to the new observations of $T_e/T_i$ in supernova remnant shocks (Rakowski 2006) and in cluster shocks (Markevitch \& Vikhlinin 2007). The SNR observations imply $1>T_e/T_i>0.1$, decreasing with shock velocity (Mach number). Although we find $T_e/T_i\approx0.5$ in the relativistic case and no significant variation in our results with shock Lorentz factor, we caution against direct comparison between our results and the nonrelativistic shocks. First, the SNR shocks may be magnetized, and here we only deal with unmagnetized shocks. Second, the nonrelativistic problem is more complicated because of the separation of scales implied by the different thermal velocities of ions and electrons. Other instabilities (such as electrostatic two-stream, firehose, etc) may also be important in nonrelativistic shocks. So, it is not surprising that the structure of these shocks may depend on more parameters. The nonrelativistic regime can, however, be explored with the existing PIC technology. 

\small{We would like to thank J. Arons, R. Blandford, Y. Lyubarsky and M. Medvedev for advice and inspiration. Simulations reported here were performed using TIGRESS computing center at Princeton University, NASA's Columbia facility and computing resources at KIPAC/SLAC. }

\references

\noindent
Birdsall, C. K. \& Langdon, A. B. 1991, ``Plasma Physics via Computer Simulations'' (McGraw-Hill: New York)

\noindent
Buneman, O. 1993 in ``Computer Space Plasma Physics'', Terra Scientific, Tokyo, 67

\noindent
Chang P., Spitkovsky A., \& Arons J. 2007, subm., arXiv:0704.3832

\noindent
Eichler, D. \& Waxman, E. 2005, \apj, 627, 861

\noindent
Frederiksen, J. T., Hededal, C. B., Haugbølle, T., \& Nordlund, Å.
2004, \apj, 608, L13

\noindent
Gruzinov, A. \& Waxman E. 1999, \apj, 511, 852

\noindent 
Hededal, C.~B., Haugbolle, T., Frederiksen, J.~T., \& Nordlund, A. 2004, \apj, 617, L107

\noindent
Hededal, C.~B. 2005, PhD thesis, astro-ph/0506559

\noindent
Lyubarsky, Y. \& Eichler, D. 2006, \apj, 647, 1250

\noindent
Markevitch, M. \& Vikhlinin A. 2007, Phys. Rep., 443, 1

\noindent
Medvedev, M.~V. \& Loeb, A. 1999, \apj, 526, 697

\noindent
Medvedev, M.~V., Fiore, M., Fonseca, R. A., Silva,
L. O.; Mori, W. B. 2005, \apj, 618, L75

\noindent
Medvedev, M.~V. 2006, \apj, 651, L9

\noindent
Milosavljevic, M., Nakar, E., \& Spitkovsky, A. 2006, \apj, 637, 765

\noindent
Nishikawa, K.-I., Hardee, P., Richardson, G., Preece, R., Sol, H., \&
Fishman, G. J. 2003, \apj, 595, 555

\noindent
Nishikawa, K.-I., Hardee, P., Richardson, G., Preece, R., Sol, H., \&
Fishman, G. J. 2005, \apj, 622, 927

\noindent
Piran, T. 2005, Rev. of Modern Phys., 76, 1143

\noindent
Rakowski, C.~E. 2006, Adv. Space Res., 35, 1017

\noindent
Silva, L. O., Fonseca, R. A., Tonge, J. W., Dawson, J. M., Mori,
W. B., \& Medvedev, M. V. 2003, \apj, 596, 121

\noindent
Spitkovsky, A. 2005, AIP Conf. Proc, 801, 345; astro-ph/0603211

\noindent
Waxman, E. 2006, Plasma Phys. Control. Fusion 48, B137

\end{document}